\documentclass[aps,prl,twocolumn]{revtex4-1}
\usepackage{amsmath,amsfonts,amssymb}
\usepackage{graphicx,float,calc}
\usepackage{color,bm}
\usepackage{ulem}
\usepackage[colorlinks,urlcolor=blue,citecolor=blue,linkcolor=blue]{hyperref}

\begin{document}

\title{Non-Hermitian Topological Anderson Insulators}

\author{Dan-Wei Zhang$^1$}\thanks{danweizhang@m.scnu.edu.cn}
\author{Ling-Zhi Tang$^1$}
\author{Li-Jun Lang$^1$}
\author{Hui Yan$^1$}
\author{Shi-Liang Zhu$^{2,1}$}\thanks{slzhu@nju.edu.cn}
\affiliation{$^1$Guangdong Provincial Key Laboratory of Quantum Engineering and Quantum Materials, GPETR Center for Quantum Precision Measurement and SPTE, South China Normal University, Guangzhou 510006, China}
\affiliation{$^2$National Laboratory of Solid State Microstructures and School of Physics, Nanjing University, Nanjing 210093, China}

\begin{abstract}

Non-Hermitian systems can exhibit exotic topological and localization properties. Here we elucidate the non-Hermitian effects on disordered topological systems using a nonreciprocal disordered Su-Schrieffer-Heeger model. We show that the non-Hermiticity can enhance the topological phase against disorders by increasing bulk gaps. Moreover, we uncover
a topological phase which emerges under both moderate non-Hermiticity and disorders, and is characterized by localized insulating bulk states with a disorder-averaged winding number and zero-energy edge modes. Such topological phases induced by the combination of non-Hermiticity and disorders are dubbed \textit{non-Hermitian topological Anderson insulators}. We reveal that the system has unique non-monotonous localization behavior and the topological transition is accompanied by an Anderson transition. These properties are general in other non-Hermitian models.

\end{abstract}

\date{\today}

\maketitle

Topological states of matter have been widely explored in condensed-matter materials \cite{Hasan2010,XLQi2011,Bansil2016,Armitage2018,Chiu2016} and various engineered systems, which include ultracold atoms \cite {DWZhang2018,Cooper2019,Goldman2016}, photonic lattices \cite{LLu2014,Ozawa2019}, mechanic systems \cite{Huber2016}, classic electronic circuits \cite{Ningyuan2015,Albert2015,Imhof2018,Lee2018b}, and superconducting circuits \cite{Schroer2014,Roushan2014,XTan2018,XTan2019a,XTan2019b}. One  hallmark of topological insulators is the robustness of nontrivial boundary states against certain types of weak disorders, since the topological band gap (topological invariants) preserves under these perturbations \cite{Hasan2010,XLQi2011,Bansil2016}. However, the band gap closes for sufficiently strong disorders and the system becomes trivial as all states are localized according to the Anderson localization \cite{Anderson1958}. Surprisingly, there is a topological phase driven from a trivial phase by disorders, known as topological Anderson insulator (TAI) \cite{JLi2009}. The TAI was first predicted in two-dimensional (2D) quantum wells and then was shown to exhibit in a wide range of systems   \cite{JLi2009,Groth2009,HJiang2009,HMGuo2010,Altland2014,Mondragon-Shem2014,Titum2015,Sriluckshmy2018,ZQZhang2019}, such as Su-Schrieffer-Heeger (SSH) chains \cite{WPSu1979}. Recently, the TAI has been observed in one-dimensional (1D) cold atomic wires and 2D photonic waveguide arrays \cite{Meier2018a,Stutzer2018}.

On the other hand, recent advances in non-Hermitian physics show that non-Hermitian systems have many intriguing features and applications \cite{Bender1998,LFeng2017,El-Ganainy2018,Miri2019}. Growing efforts are made to study topological properties of non-Hermitian systems
\cite{Rudner2009,Zeuner2015,Esaki2011,YHu2011,BZhu2014,Malzard2015,Leykam2017,YXu2017b,XZhan2017,Lee2016,SYao2018,FSong2019,Kunst2018,YXiong2018,LJin2019,Borgnia2019,ZGong2018,SYao2018b,HShen2018,Takata2018,YChen2018,LJLang2018,Harari2018,Bandres2018,HZhou2018,TSDeng2019,Ezawa2019,Kawabata2019,Ghatak2019,Kawabata2018,JQCai2018,TLiu2019,Lee2019,Yoshida2019,Luitz2019,Yamamoto2019},
which include new topological invariants \cite{Ghatak2019}, the non-Hermitian skin effect \cite{SYao2018}, the revised bulk-edge correspondence \cite{Lee2016,SYao2018,FSong2019,Kunst2018,YXiong2018,LJin2019,Borgnia2019}, and gain-and-loss induced topological phases \cite{Takata2018}. Non-Hermitian systems can exhibit unique localization properties in the presence of disorders \cite{Hatano1996,Hatano1997,Carmele2015,Mejia-Cortes2015,Levi2016,QBZeng2017,Hamazaki2018,Alvarez2018}. Notably, the topological phases have been studied in 1D non-Hermitian Aubry-Andr\'{e}-Harper model \cite{Harper1955,Aubry1980,Longhi2019,HJiang2019,QBZeng2019,JHou2019}, which describes topological quasicrystals and can be mapped to a 2D quantum Hall system \cite{Kraus2012a,LJLang2012,Ganeshan2013}. The topological non-Hermitian quasicrystals were predicted \cite{QBZeng2019,JHou2019} and the topological nature of Anderson transitions in the systems was revealed \cite{Longhi2019,HJiang2019}. However, the interplay among topology, disorder and non-Hermiticity can induce rich physical phenomena that have been rarely explored, in particular, the TAI phase has not been revealed in non-Hermitian systems.

\begin{figure}[t]
\centering
\includegraphics[width=0.45\textwidth]{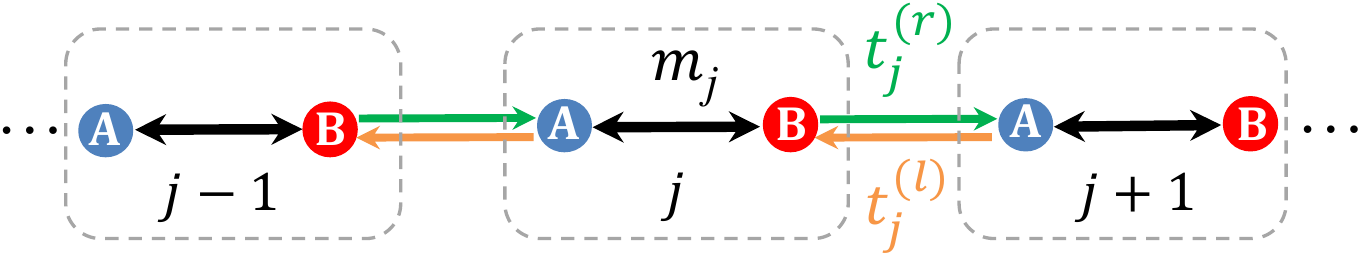}
\caption{(Color online) Sketch of the non-Hermitian disordered SSH model. The dotted box is the unit cell, and $m_j$ and $t_j^{(l,r)}$ are Hermitian intracell and nonreciprocal intercell hoppings.}
\label{fig1}
\end{figure}

In this Letter, we elucidate the non-Hermitian effects on disordered topological systems and uncover an exotic TAI phase. We construct an SSH model with nonreciprocal and disordered hopping terms and propose a real-space winding number to characterize its topology. Our main results are as follows:
(i) We show that the non-Hermiticity can enhance the topological phase against disorders under certain circumstances due to the increase of bulk gaps.
(ii) We uncover a topological phase induced by combination of non-Hermiticity and disorders, which is dubbed \textit{non-Hermitian topological Anderson insulator} (NHTAI) and characterized by localized insulating bulk states with a nontrivial winding number and zero-energy edge modes. (iii) We reveal the unique non-monotonous localization behavior owing to its interplay with the skin effect and the topological transition accompanied by an Anderson transition in the disordered system. (iv) We illustrate that the non-Hermitian enhancement effect and the NHTAI are general in other non-Hermitian models.

\textit{Model and topological invariant.---} Let us begin by considering the SSH model with nonreciprocal and disordered hoppings, which is depicted in Fig. \ref{fig1}. The tight-binding model with two-site unit cell reads
\begin{align}
H=\sum_{j}(m_ja_{j}^{\dag}b_{j}+h.c.)+t_{j}^{(r)}a_{j+1}^{\dag}b_{j}+t_{j}^{(l)}b_{j}^{\dag}a_{j+1},
\label{Ham}
\end{align}
where $a_{j}^{\dag}$ ($b_{j}^{\dag}$) creates a particle on the sublattice site A (B) in the $j$-th lattice cell, and $a_{j}$ ($b_{j}$) is the corresponding annihilation operator. Here $m_j$ denotes the $j$-dependent (Hermitian) intracell hopping energy, and $t_j^{(r,l)}$ characterize the non-Hermitian intercell hoppings. This Hamiltonian has the chiral symmetry as $H$ satisfies $CHC^{-1}=-H$, where the chiral operator is $C=\sigma_z\otimes \mathbb{I}$ with the Pauli matrix $\sigma_z$ referring to the sublattice and the identity matrix $\mathbb{I}$ acting on the lattice cell.

In contrast to the site-potential disorder, the pure tunneling disorder is crucial for preserving the chiral symmetry. In particular, we consider the hopping terms as
\begin{equation}
m_j=t+W_1\omega_{j}, ~t_j^{(l)}=t'+W_2\omega'_{j}, ~t_j^{(r)}=t_j^{(l)}+f(\gamma).
\end{equation}
Here $t$ and $t'$ are intracell and intercell tunneling energies, $\omega_{j}$ and $\omega'_{j}$ are independent random numbers chosen uniformly in the range $[-1,1]$, $W_1$ and $W_2$ are the disorder strengths. Unless mentioned otherwise, we hereafter assume the nonreciprocal term $f(\gamma)=t'\gamma$ with the non-Hermiticity parameter $\gamma$. We set $t=1$ as the energy unit and focus on the case of $W_2=0$.

Similar as the topological phase in the SSH model with $t'>t$, this non-Hermitian disordered SSH model is topologically characterized by zero-energy edge modes and the corresponding winding number [see Eq. (\ref{windnum2})]. Note that the system recovers to the Hermitian disordered SSH chain when $\gamma=0$ \cite{Meier2018a,Altland2014,Mondragon-Shem2014} and to the non-Hermitian clean chain when $W_1=W_2=0$ \cite{Lee2016,SYao2018,FSong2019,Kunst2018}, respectively. In the clean limit, the topological invariant of the nonreciprocal SSH model can be a non-Bloch winding number in complex momentum space \cite{SYao2018} or a dual open-bulk winding number in real space \cite{FSong2019}.

We now generalize the open-bulk winding number to our non-Hermitian disordered SSH model. Given a disorder configuration denoted by $s$, we diagonalize the Hamiltonian (\ref{Ham}) under open boundary conditions (OBCs) with two chiral-symmetric parts: $H_s|nR_{\pm}\rangle_s=\pm E_{n,s}|nR_{\pm}\rangle_s$ with $|nR_-\rangle_s=C|nR_+\rangle_s$. In the biorthonormal basis, the corresponding left eigenstates $|nL_{\pm}\rangle$ orthonormal to the right eigenstates
can be taken from the columns of $(T_s^{-1})^\dag$ by writing $H_s=T_s\Lambda_s T_s^{-1}$ with $\Lambda_s$ diagonal. The homotopically equivalent flat band version of the Hamiltonian $H_s$ under OBCs is the open-boundary $Q_s$ matrix, which is given by $Q_s=\sum_{n} \left(|nR_+\rangle_s{_s}\langle nL_+|-|nR_-\rangle_s{_s}\langle nL_-|\right)$. Here the summing takes over the eigenstates in the bulk spectrum without edge modes. The open-bulk winding number in real space is then defined as
\begin{equation}
\nu_s= \frac{1}{2L'}\text{Tr}' (C Q_s[Q_s,X]), \label{windnum1}
\end{equation}
where $X$ is the coordinate operator, $L=L'+2l$ is the chain length with three intervals of lengthes $l, L', l$, and $\text{Tr}'$ denotes the trace over the middle interval of length $L'$  \cite{FSong2019}.
Here $\nu_s$ serves well for disordered systems as it does not require the translation invariance and is quantized to an integer in the limit of $L\rightarrow\infty$ \cite{Kitaev2006a,Mondragon-Shem2014}. We can define the disorder-averaged winding number
\begin{equation}
\nu=\frac{1}{N_s}\sum_{s=1}^{N_s}\nu_s, \label{windnum2}
\end{equation}
where a modest configuration number $N_s$ suffices in practice. For $\gamma=0$, the topological invariants in Eqs. (\ref{windnum1}) and (\ref{windnum2}) reduce to those in Hermitian systems \cite{Kitaev2006a,Prodan2010,Bianco2011,YXZhao2017,Meier2018a,Mondragon-Shem2014}, where the boundary condition is irrelevant. However, Eq. (\ref{windnum1}) correctly corresponds to the topological edge modes only under OBCs, due to the unconventional bulk-edge correspondence in the non-Hermitian cases \cite{Lee2016,SYao2018,FSong2019,Kunst2018,YXiong2018,LJin2019}.

\begin{figure}[t]
\centering
\includegraphics[width=0.45\textwidth]{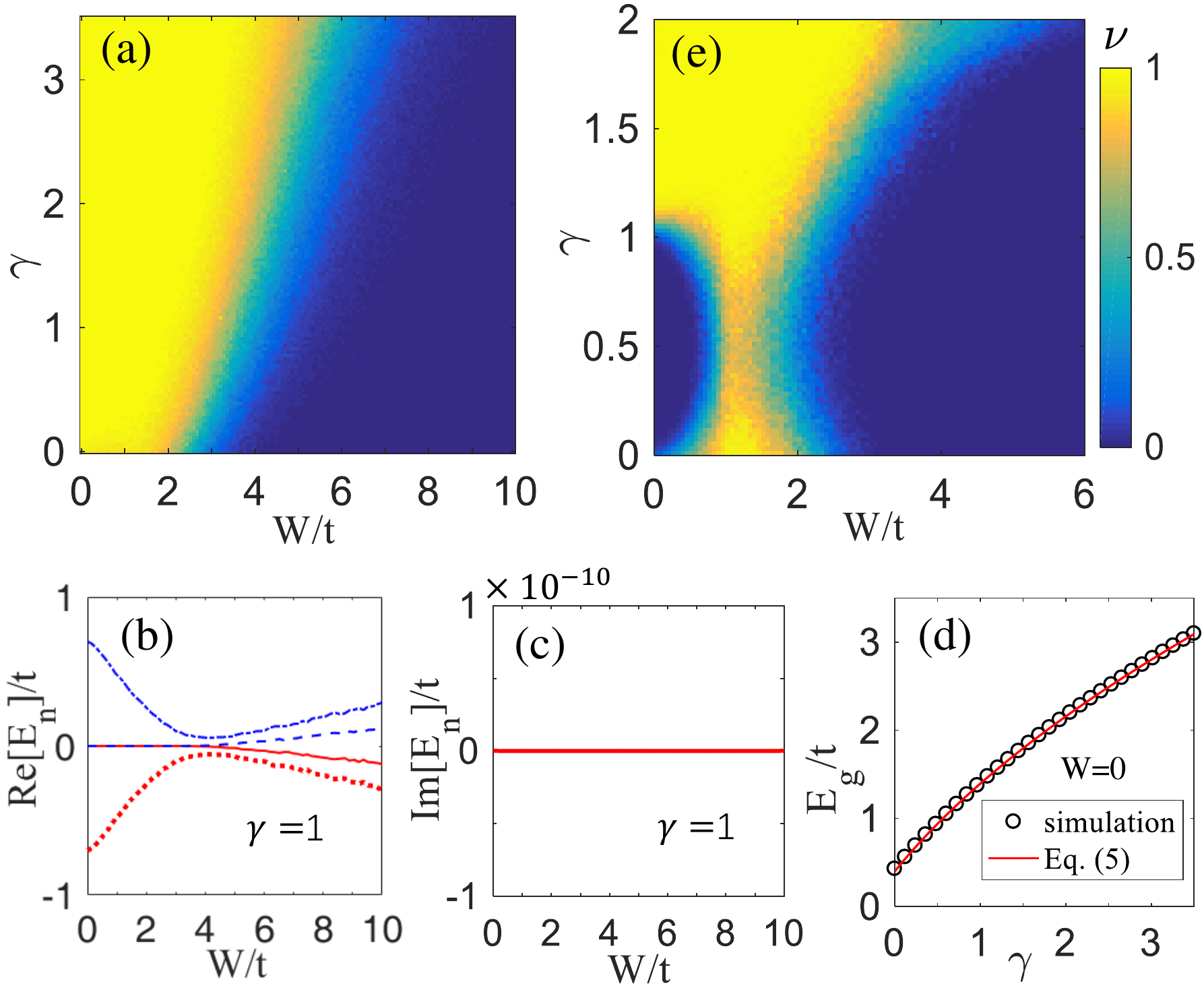}
\caption{(Color online) 
(a) Disorder-averaged winding number $\nu$ as a function of non-Hermiticity $\gamma$ and disorder strength $W$. (b,c) Four middle disorder-averaged energy eigenvalues $E_n$ (real and imaginary parts) under OBCs, as a function of $W$ for $\gamma=1$. From bottom to up are $n=L/2-1,L/2,L/2+1,L/2+2$, respectively. (d) The bulk gap $E_g$ as a function of $\gamma$ from simulations (hollow dots) and Eq. (\ref{Egap}) (sold line). (e) $\nu$ as a function of $\gamma$ and $W$ for a nonreciprocal term $f(\gamma)=t'(1-\gamma+\gamma^2)$ with $t'=t$. Other parameters are $t=1$, $t'=1.2t$ [in (a-d)], $L=5l=100$, $W_1=W$, $W_2=0$ and $N_s=200$.}
\label{fig2}
\end{figure}

\textit{Enhancing topological phase by non-Hermiticity.---}We first consider the non-Hermitian effect on the topological phase of the open SSH chain with $t'>t$ under disorders. Figure \ref{fig2}(a) shows the disorder-averaged winding number $\nu$ as a function of $\gamma$ and disorder strengths $W_1=W$ and $W_2=0$. We can find that when $\gamma$ varies from $0$ to $3.5$, the topological phase with $\nu\simeq1$ preserves from a region $W/t\lesssim2$ to a larger one $W/t\lesssim5$.
We rewrite the eigenequation of the open chain $H_s|\psi_n\rangle_s=E_n^{(s)}|\psi_n\rangle_s$ with wave functions $|\psi_n\rangle_s=[\psi_{n,1}^{(s)},\psi_{n,2}^{(s)},...\psi_{n,x}^{(s)},...\psi_{n,L}^{(s)}]^T$ and eigenenergies $E_n^{(s)}$, where $x$ is the lattice site index. Then the disorder-averaged eigenenergies is given by $E_n=\frac{1}{N_s}\sum_{s=1}^{N_s}E_n^{(s)}$. Figures \ref{fig2}(b,c) show the real and imaginary parts of $E_n$ for four center eigenstates as a function of $W$, respectively. The energy spectrum of this open chain is purely real and two zero-energy edge modes exhibit in the topological phase.

To understand the numerical results, we take a similarity transformation \cite{SYao2018}: $\tilde{H}_s=S^{-1}H_sS$ with the diagonal matrix $S=\text{diag}(1,1,r,r,r^2,r^2,...,r^{L/2-1},r^{L/2-1})$, the eigenequation is equivalent to $\tilde{H}_s|\tilde{\psi}_n\rangle_s=E_n^{(s)}|\tilde{\psi_n}\rangle_s$ with $|\tilde{\psi}_n\rangle_s=S^{-1}|\psi_n\rangle_s$. Let $r=\sqrt{1+\gamma}$ for $\gamma>-1$, then $\tilde{H}$ becomes the Hermitian disordered SSH model with intracell and
intercell hoppings $\tilde{m}_j=m_j$ and $\tilde{t}'=t'\sqrt{1+\gamma}$ for $W_2=0$. This transformation indicates that all eigenenergies of the open-chain Hamiltonian are real [Figs. \ref{fig2}(b,c)]. The transformation also accumulates the wave functions of bulk states to one boundary, which is the non-Hermitian skin effect \cite{SYao2018} and will be discussed later. We can numerically calculate the bulk gap of the open clean system $E_g=|E_{L/2+2}-E_{L/2-1}|$ for $W=0$. Furthermore,  the bulk gap for the open non-Hermitian chain can be derived from the Hermitian SSH chain after  the similarity transformation, i.e.,
\begin{equation}\label{Egap}
\tilde{E}_g=2|\tilde{t}'-\tilde{t}|=2|t'\sqrt{1+\gamma}-t|.
\end{equation}
The numerical results of $E_g$ for $L=100$ are consistent with the values of $\tilde{E}_g$ given by Eq. (\ref{Egap}), as shown in Fig. \ref{fig2}(d). Thus, the enhancement of the topological phase in this case can be interpreted as the increase of the effective bulk gap by non-Hermiticity.

The enhancement effect would be broken for negative $\gamma$ or large $W_2$, however, it can preserve without a similarity transformation, such as in the cases with small $W_2$ or an additional intercell hopping \cite{Note1}. Surprisingly, we find that the nonreciprocal hopping (the gain and loss) can induce a topological phase when the corresponding Hermitian limit of $\gamma=0$ is a trivial or critical phase for $t' \leq t$. The results for the critical case of $t=t'$ are shown in Fig. \ref{fig2}(e) for $f(\gamma)=t(1-\gamma+\gamma^2)$ (in Fig. \ref{figS5} for the disordered SSH model with non-Hermitian gain and loss \cite{Note1}). From Fig. \ref{fig2}(e), we also find the combination of non-Hermiticity and disorders can give rise to the topological phase.

\begin{figure}[t]
\centering
\includegraphics[width=0.48\textwidth]{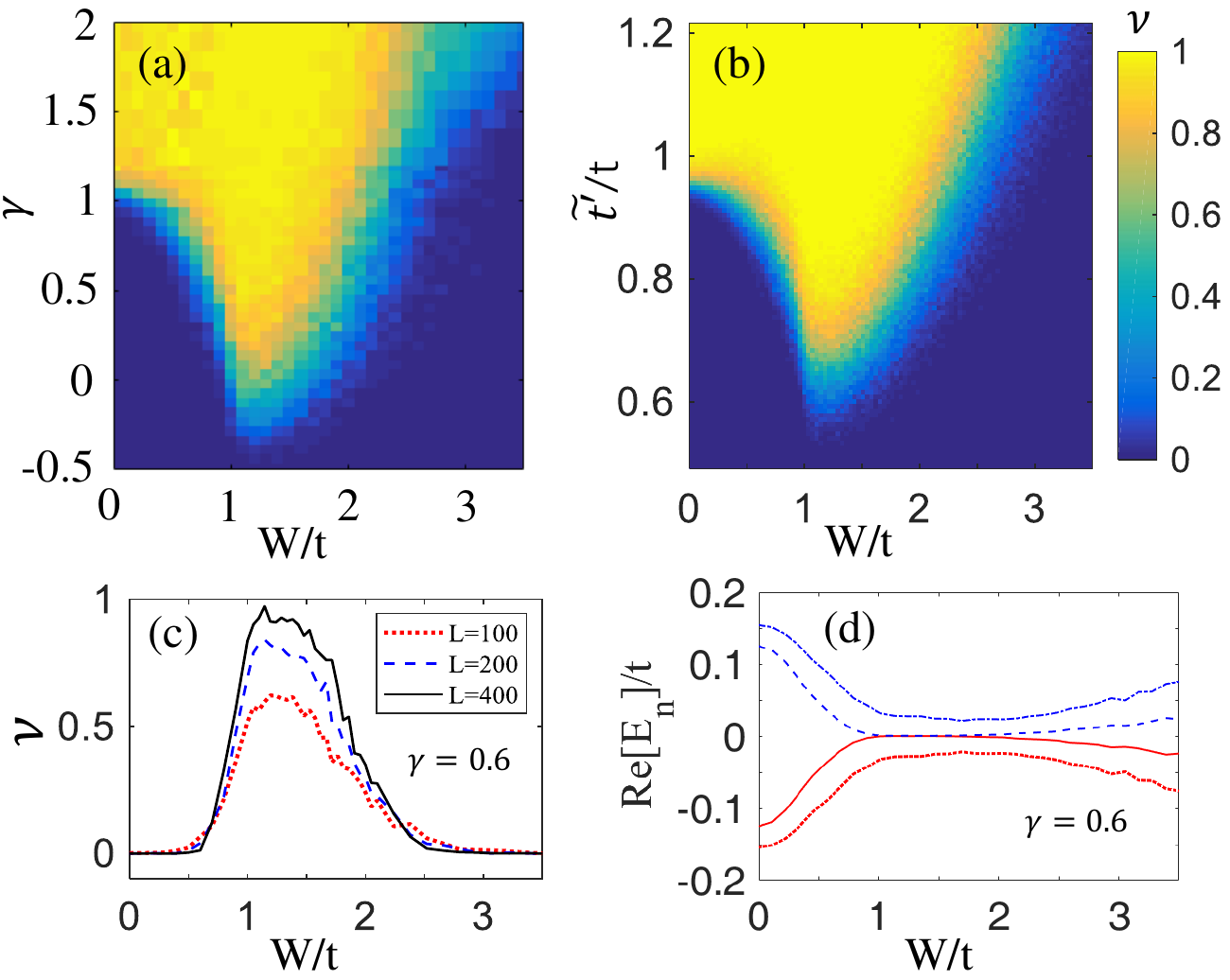}
\caption{(Color online) 
The winding number $\nu$ as a function of $W$ and $\gamma$ (a) or $\tilde{t}'$ (b) for $L=400$. (c) $\nu$ as a function of $W$ for fixed $\gamma=0.6$ and $L=100,200,400$. (d) Four disorder-averaged eigenenergies in the center of the energy spectrum $E_n=\text{Re}[E_n]$ under OBCs, as a function of $W$ for $\gamma=0.6$ and $L=100$. From bottom to up are $n=L/2-1,L/2,L/2+1,L/2+2$, respectively. Other parameters are $t'=0.7t$, $l=0.2L$, $W_1=W$, $W_2=0$, and $N_s=50$.}
\label{fig3}
\end{figure}

\textit{NHTAI from non-Hermiticity and disorder.---}We further study the effects of combined non-Hermiticity and disorder on an initially trivial phase in the SSH chain with $t>t'$. We numerically calculate $\nu$ as a function of $\gamma$ and disorder strengths $W_1=W$ and $W_2=0$, with the results for $t'=0.7t$ and $L=400$ shown in Fig. \ref{fig3}(a). In the Hermitian and clean limit $\gamma=W=0$, the system is in the trivial phase with $\nu\simeq0$. Interestingly, we find $\nu\approx1$ in a region with moderate non-Hermiticity and disorder strength in Fig. \ref{fig3}(a). Actually, $\nu$ can approach to unit in this region by increasing the lattice size, with an example as a function of $W$ with $\gamma=0.6$ shown in Fig. \ref{fig3}(c). The four middle disorder-averaged eigenenergies of the open chain of $L=100$ are plotted in Fig. \ref{fig3}(d), which shows that two zero-energy edge modes inside a small gap between upper and lower eigenstates corresponding to the topological phase. Note that the small gap essentially vanish when $L\rightarrow\infty$ 
for moderate disorders. It is excepted that the energy gap is replaced by a mobility gap and the band insulator becomes an Anderson insulator with the topology carried by localized bulk states, similar as those for the Hermitian disordered chiral wires \cite{Mondragon-Shem2014}. Our results indicate that a topological insulator can be induced from a trivial phase in the Hermitian and clean limit by the combination of moderate non-Hermiticity and disorders, which is dubbed NHTAI as a non-Hermitian extension of the TAI \cite{JLi2009,Groth2009,HJiang2009,HMGuo2010,Meier2018a,Altland2014,Mondragon-Shem2014}.

To show the connection between NHTAIs and TAIs, we perform the similarity transformation and map the non-Hermitian open SSH chain to the Hermitian one with $\tilde{t}'=t'\sqrt{1+\gamma}$. 
We calculate the corresponding winding number for the Hermitian Hamiltonian (where the right eigenstates are used) \cite{Mondragon-Shem2014,Meier2018a}. The results for an open chain of $L=400$ shown in Fig. \ref{fig3}(b) indicates the $\tilde{t}'$-$W$ region of the TAI for $0.7<\tilde{t}'/t<0.97$, which corresponds well to the $\gamma$-$W$ region of the NHTAI for $0<\gamma<0.95$. Thus, NHTAIs can be topologically connected to TAIs through the similarity transformation with the same energy spectrum under OBCs. Notably, the NHTAI has unique properties without Hermitian counterparts, such as unconventional bulk-edge correspondence and non-monotonous localization behavior under OBCs as we will discuss below. Moreover, the NHTAI can exist without the similarity transformation when the energy spectrum is generally complex \cite{Note1}.

\begin{figure}[t]
\centering
\includegraphics[width=0.46\textwidth]{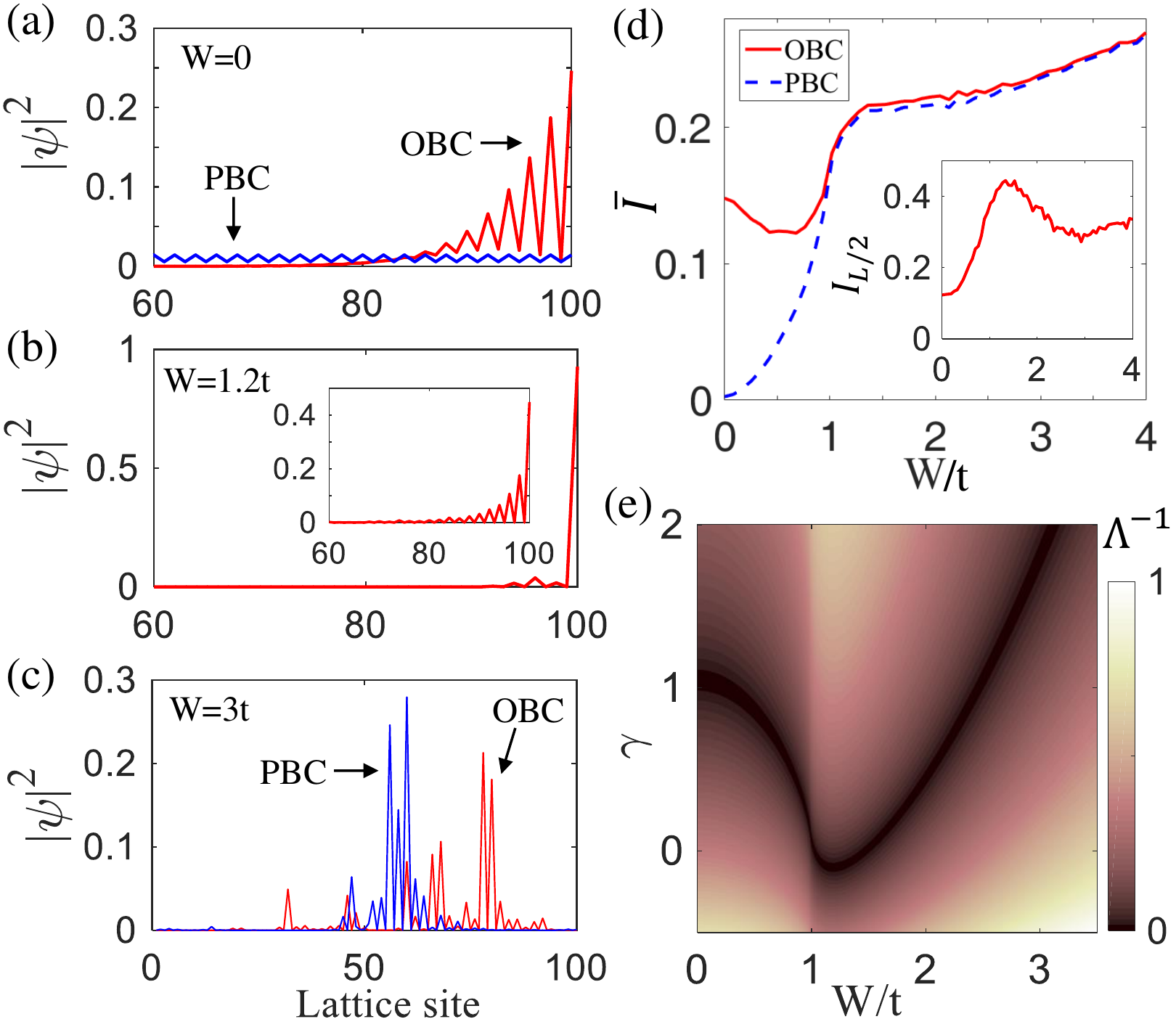}
\caption{(Color online) (a,b,c) Density distribution of the $L/2$-th eigenstate with $L=100$ for $W/t=0,1.2,3$ in a disorder configuration, respectively. The inset figure in (b) is the disorder-averaged density distribution. (d) $\bar{\mathcal{I}}$ and $\mathcal{I}_{L/2}$ (inset) with $L=400$, respectively. (e) $\Lambda^{-1}$ as a function of $\gamma$ and $W$. The same parameters as in Fig. \ref{fig3} and $\gamma=0.6$ in (a-d).}
\label{fig4}
\end{figure}

\textit{Localization properties.---}We now investigate the localization properties of the disordered system. The density distributions of the $L/2$-th eigenstate ($L=100$) for $\gamma=0.6$ and $W/t=0,1.2,3$ in a disorder configuration are shown in Figs. \ref{fig4}(a-c), respectively. In the clean limit [Fig. \ref{fig4}(a)], this eigenstate is a bulk state, which is extended under PBCs but pinned to the right edge (because $t_j^{(r)}>t_j^{(l)}$) of the lattice under OBCs, as a manifestation of the non-Hermitian skin effect \cite{SYao2018}. For moderate disorder strength, this eigenstate becomes a zero-energy mode localized at the right edge of the system that is in the NHTAI phase. The disorder-averaged density distribution of the edge state is shown in the inset of Fig. \ref{fig4}(b). For large disorder strength [Fig. \ref{fig4}(c)], it becomes a localized state in the bulk and the skin effect is broken. The same results are for the $(L/2+1)$-th eigenstate due to the chiral symmetry. We calculate the disorder-averaged inverse participation ratio for the $n$-th eigenstate $\mathcal{I}_n$ and its averaging over all eigenstates $\bar{\mathcal{I}}$, which are given by
\begin{equation}
\mathcal{I}_n=\frac{1}{N_s}\sum_{s=1}^{N_s}\sum_{x=1}^{L}|\psi_{n,x}^{(s)}|^4,~~\bar{\mathcal{I}}=\frac{1}{L}\sum_{n=1}^{L}\mathcal{I}_n,
\end{equation}
respectively. We find that $\mathcal{I}_n>1/L$ for all bulk states with $E_n\neq0$ when $W>0$, which implies that the entire energy spectrum (excluded $E=0$) of the chiral chain are localized immediately after the disorder is turned on \cite{Mondragon-Shem2014}. The results for $\bar{\mathcal{I}}$ and $\mathcal{I}_{L/2}$ with $L=400$ are shown in Fig. \ref{fig4}(d). Under OBCs, $\mathcal{I}_{L/2}\gg1/L$ for all $W$ indicates that the $L/2$-th eigenstate always localized in this case; and $\mathcal{I}_{L/2}$ with a maximum value in the NHTAI phase implies that the zero-energy edge modes are more localized. The global localization index $\bar{\mathcal{I}}$ increasing rapidly as a function of $W$ under PBCs also shows that the bulk states are localized in the presence of disorders. Remarkably, $\bar{\mathcal{I}}$ varies non-monotonously when $W/t\lesssim1$ under OBCs, owing to the interplay of Anderson localization and the skin effect. When $W/t\gtrsim1$, $\bar{\mathcal{I}}$ takes nearly the same values for OBCs and PBCs, which indicates that the skin effect of bulk states is now destroyed by disorders.

Following Ref. \cite{Mondragon-Shem2014}, we derive an analytical formula for the localization
length $\Lambda$ of our model (with $W_2=0$) at energy $E=0$ under OBCs. We obtain \cite{Note1}
\begin{equation}
\Lambda^{-1}=\left|\ln\left[\frac{2et'\sqrt{1+\gamma}  |2-2W|^{\frac{1}{2W}-\frac{1}{2}}}{|2+2W|^{\frac{1}{2W}+\frac{1}{2}}}  \right]\right|,
\end{equation}
where $e$ is the natural constant. Figure. \ref{fig4}(e) shows the critical points with $\Lambda^{-1}\rightarrow0$ in the $\gamma$-$W$ plane, where the localization length of zero-energy states diverges. The delocalized critical points
match with the topological transition points in Fig. \ref{fig3}(a), indicating that the topological transition is accompanied by an Anderson localization-delocalization transition. Our results reveal that the non-Hermitian topological numbers
can be carried by localized bulk states and disorders can drive a localized non-Hermitian topological phase through a delocalized point, similar as those in Hermitian chiral chains \cite{Mondragon-Shem2014}.

\textit{Discussion and conclusion.---}We discuss the realization of the non-Hermitian disordered SSH model in some artificial systems. The first feasible system is ultracold atoms \cite{DWZhang2018,Goldman2016,Cooper2019}. The TAI has been realized in an atomic SSH wire with controllable disordered hoppings \cite{Meier2018a}, where the effective nonreciprocal hopping can be engineered by a collective one-body loss \cite{ZGong2018,JLi2019}. Another system is photonic crystals \cite{LLu2014,Ozawa2019}, where the TAI phase and tunable non-Hermiticity were achieved \cite{Stutzer2018,Poli2015,Weimann2016,MPan2018,Oezdemir2019}. The realization of nonreciprocal hoppings in optics was suggested \cite{Longhi2014,Longhi2015b}. The third system is topological electronic circuits \cite{Ningyuan2015,Albert2015,Imhof2018,Lee2018b}, where the Hermitian and nonreciprocal SSH chains were realized \cite{Hadad2018,YWang2019,Helbig2019} and tunable hopping disorders can be added. In view that non-Hermiticity and disorders have been engineered in these artificial systems, the studied model with the NHTAI phase is realizable in current experiments. The robust or disorder-induced topological edge modes can be detected, and it would be interesting to measure the topological numbers.

We note that our findings are applicable for other non-Hermitian models. The non-Hermitian enhancement effect and the NHTAI phase can preserve without the similarity transformation, in which cases the energy spectra are generally complex. For instance, when $W_2\neq0$, the similarity transformation is inapplicable but the main results obtained for $W_2=0$ preserve for $W_2\lesssim W/4$, as shown in Supplemental Material \cite{Note1}. Another example is a modified model with an additional intercell hopping term \cite{Note1}. The enhancement effect and the NHTAI phase also exhibit in a disordered SSH model with non-Hermitian gain and loss \cite{Note1}. The NHTAI phase even exists when the non-Hermiticity becomes random, although the skin effect is generally broken in this case \cite{Note1}.

In summary, we have explored the topological and localization properties of the SSH model with nonreciprocal and disordered hoppings. We have revealed the enhancement of the topological phase by non-Hermiticity and the NHTAI induced by combination of non-Hermiticity and disorders. The non-monotonous localization behavior and the topological nature of the Anderson transition have been elucidated. Moreover, the predicted NHTAI can be experimentally realized in some artificial systems.

\begin{acknowledgments}
This work was supported by the NKRDP of China (Grant No. 2016YFA0301800), the NSFC (Grants No. 11604103, No. 11704367, No. 11904109, and No. 91636218), the NSAF (Grant No. U1830111 and No. U1801661),  the Key-Area Research and Development Program of GuangDong Province (Grant No. 2019B030330001), and the Key Program of Science and Technology of Guangzhou (Grant No. 201804020055).
\end{acknowledgments}

\bibliography{reference}

\clearpage

\appendix

\section{Supplemental Material for \\ \textit{Non-Hermitian Topological Anderson Insulators}}

\subsection{1. Localization length at zero energy}

For the Hermitian disordered chiral SSH chain, the localization length $\Lambda$ at energy $E=0$ can be analytically obtained \cite{Mondragon-Shem2014}. Indeed, the Schr\"{o}dinger equation of the Hermitian SSH model $\tilde{H}\tilde{\psi} =0$ reads: $\tilde{t}_j\tilde{\psi}_{j-\alpha,\alpha}+ \tilde{m}_j \tilde{\psi}_{j,\alpha} =0$, where $\alpha = \pm1$ represents A and B sublattice, respectively. The solution is given by
\begin{equation} \label{wavefunction-E0}
\tilde{\psi}_{N,\alpha}=\prod_{j=1}^{N} \left( \frac{\tilde{t}_j}{\tilde{m}_j}\right ) \tilde{\psi}_{0,\alpha},
\end{equation}
where the unit cell is labeled by $j=0,1,...,N$. The inverse of the localization length is given by (in the thermodynamic limit disregarding the boundaries) \cite{Mondragon-Shem2014}
\begin{equation}\label{local-length}
\begin{array}{l}
\Lambda^{-1} =\max_{\alpha=\pm 1}\big [- \lim\limits_{N\rightarrow \infty} \frac{1}{N}\ln|\tilde{\psi}_{N,\alpha}| \big ] \medskip \\
\indent =\big |\lim\limits_{N\rightarrow \infty} \frac{1}{N}\sum_{j=1}^{N}(\ln |\tilde{t}_j|-\ln|\tilde{m}_j|)\big |.
\end{array}
\end{equation}
We consider the disordered hopping parameters: $\tilde{m}_j=\tilde{t}+W_1\omega_{j}$ and $\tilde{t}_j=\tilde{t}'+W_2\omega'_{j}$, where $t$ and $t'$ are the characteristic intracell and intercell tunneling energies, $\omega_{j}$ and $\omega'_{j}$ are independent random real numbers chosen uniformly from the range $[-1,1]$, $W_1$ and $W_2$ are the corresponding disorder strengths, and $\tilde{t}=1$ is set as the energy unit. Note that the notations used here are slightly different from those in Ref. \cite{Mondragon-Shem2014}.

According to Birkhoff's ergodic theorem, one can use the ensemble average to evaluate the last expression in Eq. (\ref{local-length}), which is then given by
\begin{equation}
\begin{array}{l}
\Lambda^{-1} =\left | \int_{-1}^{1}d\omega \int_{-1}^{1}d\omega' \ (\ln|\tilde{t}'+W_2 \omega| - \ln|1+W_1 \omega'|)\right |.
\end{array}
\end{equation}
The integrations can be calculated explicitly, and the arguments of the logarithms can become negative in the regime of large $W$. One can obtain
\begin{equation}\label{LocLength}
\Lambda^{-1}=\left | \ln \left [\frac{|2\tilde{t}'+2W_2|^{\frac{\tilde{t}'}{2W_2}+\frac{1}{2}}}{|2\tilde{t}'-2W_2|^{\frac{\tilde{t}'}{2W_2}-\frac{1}{2}}}  \frac{| 2-2W_1|^{\frac{1}{2W_1}-\frac{1}{2}}}{|2+2W_1|^{\frac{1}{2W_1}+\frac{1}{2}}}  \right ] \right |.
\end{equation}

\begin{figure}[t]
\centering
\includegraphics[width=0.35\textwidth]{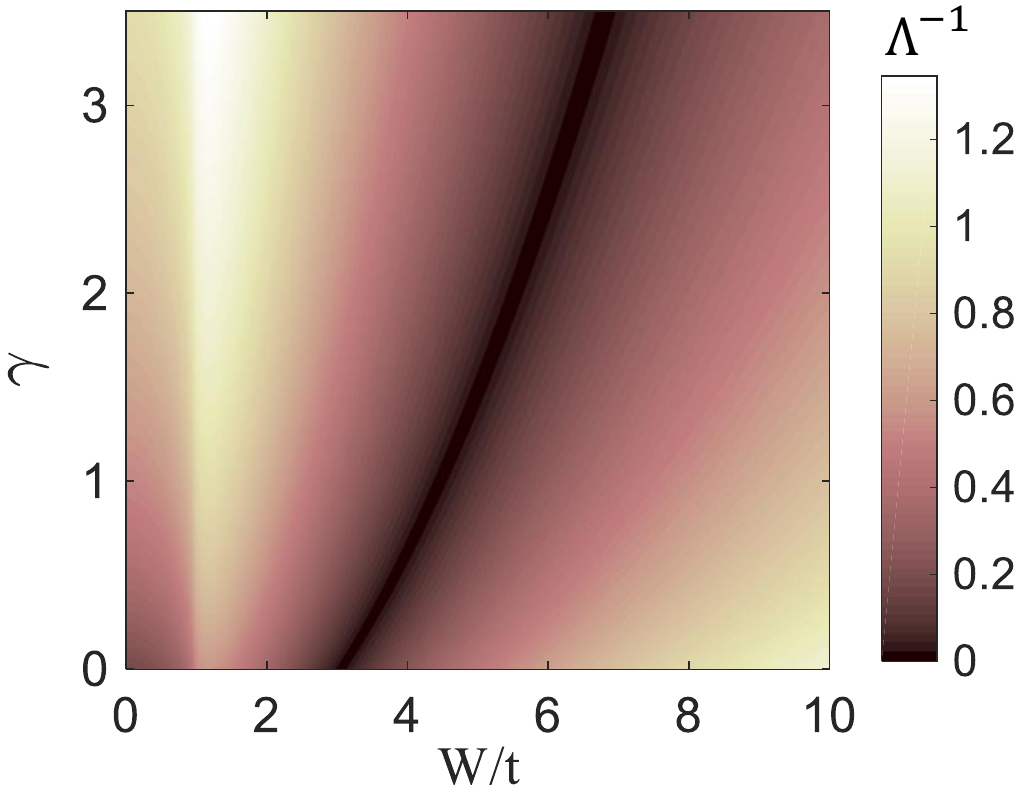}
\caption{(Color online) $\Lambda^{-1}$ as a function of $\gamma$ and $W$. The parameters are $t=1$, $t'=1.2t$, $W_1=W$, $W_2=0$.}
\label{figS1}
\end{figure}

By using the similarity transformation $\tilde{H}=S^{-1}HS$ on our non-Hermitian disordered SSH Hamiltonian $H$ with $W_2=0$ under OBCs, one can the Hermitian disordered SSH Hamiltonian $\tilde{H}$. After the transformation, we can obtain the solution of the Schr\"{o}dinger equation of $H\psi=0$ with $\psi=S\tilde{\psi}$, which replaces the solution in Eq. (\ref{wavefunction-E0}) with the form:
\begin{equation}
\psi_{N,\alpha}=\prod_{j=1}^{N} \left( \frac{t_j}{m_j}\right ) \psi_{0,\alpha}.
\end{equation}
With the mapping, one has the hopping parameters $t_j=t'r+W_1\omega_{j}=t'\sqrt{1+\gamma}+W_1\omega_{j}$ and $m_j=\tilde{m}_j=1$ for $\tilde{t}=t=1$ and $W_2=0$. Thus, by substituting $\tilde{t}'=t'\sqrt{1+\gamma}$, $W_1=W$ and $W_2=0$ into Eq. (\ref{LocLength}), we obtain the inverse of the localization length of zero-energy states in our non-Hermitian disordered SSH chain under OBCs,
\begin{equation}
\Lambda^{-1}=\left|\ln\left[\frac{2et'\sqrt{1+\gamma}  |2-2W|^{\frac{1}{2W}-\frac{1}{2}}}{|2+2W|^{\frac{1}{2W}+\frac{1}{2}}}  \right]\right|,
\end{equation}
where $e$ is the natural constant. Figure. \ref{figS1} shows the critical points with $\Lambda^{-1}\rightarrow0$ in the $\gamma$-$W$ plane, where the delocalized critical points match with the topological transition points in Fig. 2(a). This result, as well as Fig. 4(e) and Fig. 3(a) in the main text, demonstrates that the topological transition is accompanied by an Anderson localization-delocalization transition in our non-Hermitian disordered SSH model.

\begin{figure*}[t]
\centering
\includegraphics[width=0.95\textwidth]{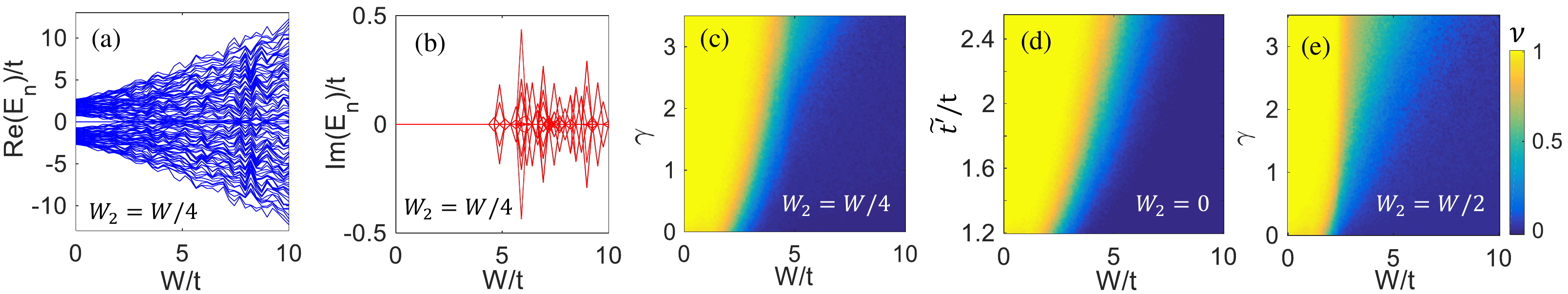}
\caption{(Color online)
(a,b) The energy spectrum $E_n$ (real and imaginary parts) under OBCs as a function of $W$ with fixed $\gamma=1$ in a disorder configuration with strengths $W_1=4W_2=W$. (c,e) Disorder-averaged winding number $\nu$ as a function of $W$ ($W_2=W/4$ and $W_2=W/2$) and $\gamma$; (d) $\nu$ as a function of $W$ ($W_2=0$) and $\tilde{t}'$. Other parameters are $t=1$, $t'=1.2t$, $L=5l=100$, and $N_s=200$.}
\label{figS2}
\end{figure*}

\begin{figure}[t]
\centering
\includegraphics[width=0.42\textwidth]{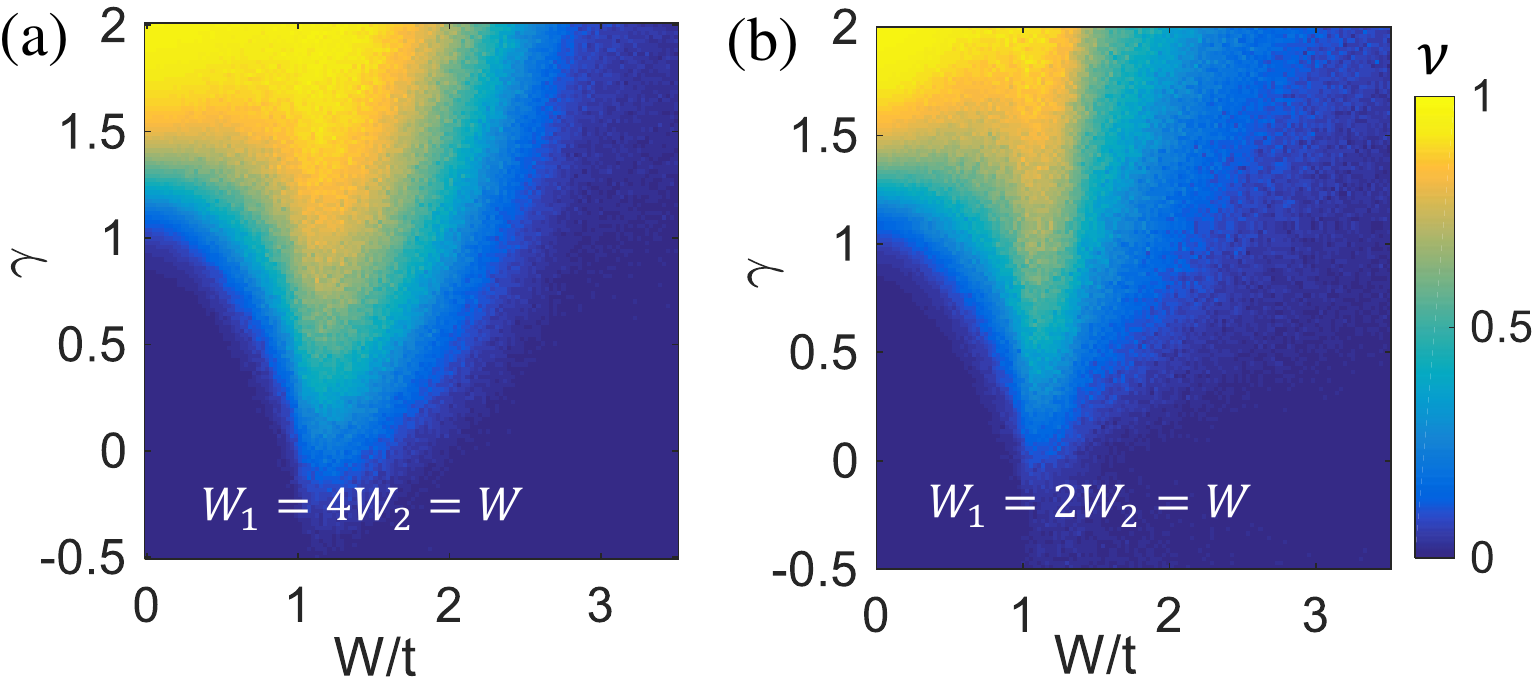}
\caption{(Color online) (a,b) $\nu$ as a function of $W$ and $\gamma$. Other parameters are $t=1$, $t'=0.7t$, $l=0.2L$, and $N_s=200$.}
\label{figS3}
\end{figure}

\subsection{2. Results for the case of $W_2\neq0$}

When the intercell hopping disorder strength $W_2\neq0$, one can not find a similarity transformation of the non-Hermitian disordered SSH chain under OBCs for every disorder configuration (especially in the strong disorder case). In this case, the (disorder-averaged) energy spectrum can be generally complex. Figures. \ref{figS2}(a,b) show the complex energy spectrum for $W_1=4W_2=W$ as a function of $W$ in a disorder configuration. The corresponding disorder-averaged winding number $\nu$ as a function of $W$ and $\gamma$ is shown in Fig. \ref{figS2}(c), which is close to the result of $W_2=0$ shown in Fig. 2(a). Thus, the non-Hermitian enhancement of the topological phase can still exhibit when $W_2\neq0$. For comparisons, we plot the winding number $\nu$ as a function of $W$ and $\tilde{t}'$ in Fig. \ref{figS2}(d), corresponding to the Hermitian open SSH chain with $\tilde{t}'=t'\sqrt{1+\gamma}$ after the similarity transformation when $W_2=0$, where the topological regime matches well with those in Figs. 2(a) and \ref{figS2}(c). We note that the enhancement effect will be broken by increasing $W_2$, such as when $W_2\gtrsim W/2$ in this case, as shown in Fig. \ref{figS2}(e).

In Figs. \ref{figS3}(a,b), we show $\nu$ as a function of $W$ and $\gamma$ for $W_1=4W_2=W$ and $W_1=2W_2=W$, respectively. Although we here adopt a lattice of length $L=100$ smaller than that of $L=400$ in Fig. 3, the topological regime with the NHTAI can be roughly seen. Tuning on the disorder of $W_1$ up to $W_1=W/4$, the topological regime almost preserves, and the NHTAI phase remains. However, the topological region will be reduced when $W_1$ becomes larger, such as the case of $W_1=W/2$ in Fig. \ref{figS3}(b) with a narrow parameter region for the NHTAI phase.

\subsection{3. Modified Hamiltonian without a similarity transformation}

\begin{figure}[t]
\centering
\includegraphics[width=0.45\textwidth]{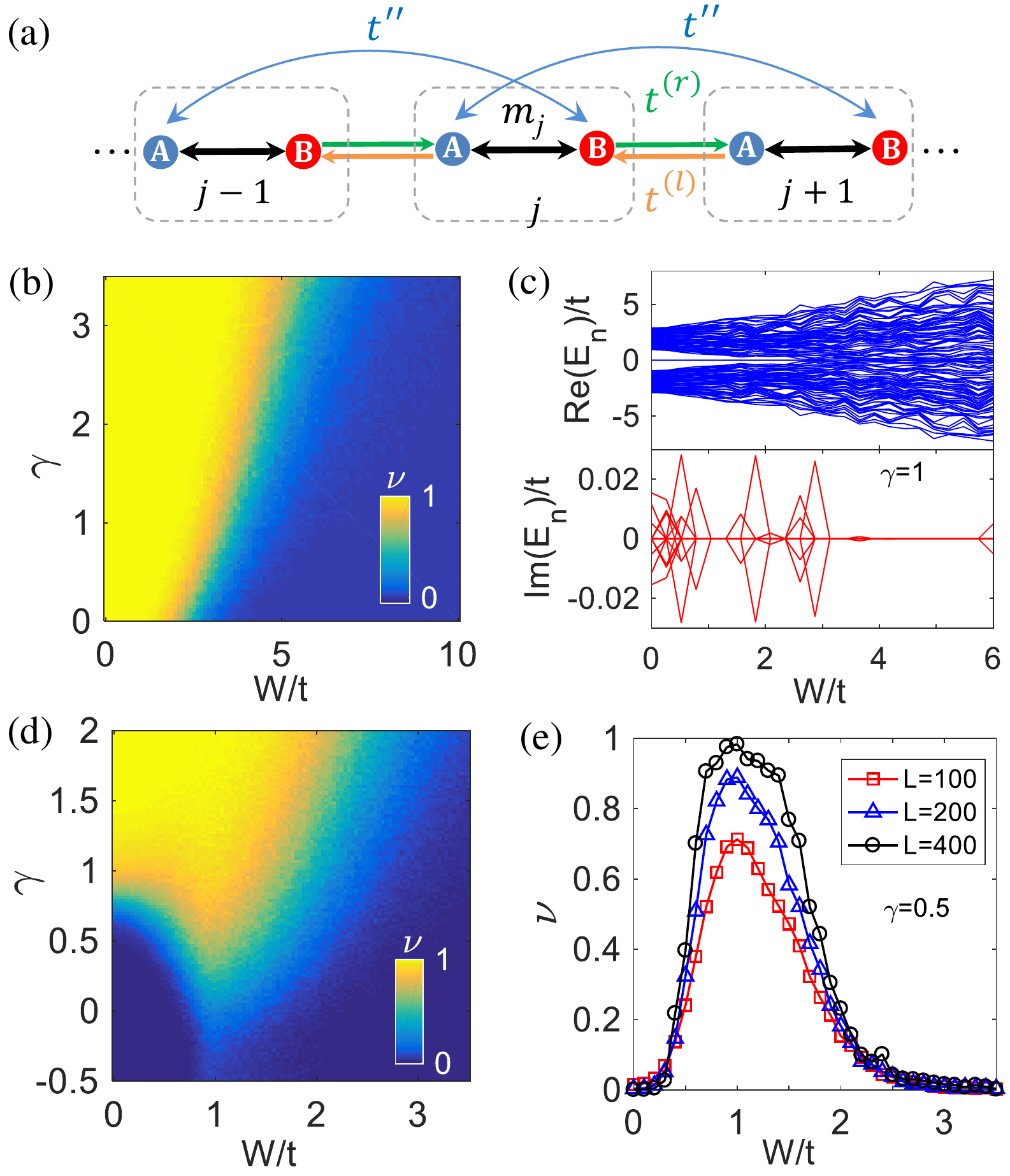}
\caption{(Color online) (a) Sketch of the disordered nonreciprocal SSH model with an additional hopping term $t''$. (b,d) $\nu$ as a function of $W$ and $\gamma$ for $t''=0.2t$ and $t''=0.1t$, respectively. (c) The energy spectrum $E_n$ (real and imaginary parts) under OBCs as a function of $W$ for $\gamma=1$ in a disorder configuration. (e) $\nu$ as a function of $W$ for fixed $\gamma=0.5$, $t''=0.1t$, and $L=100,200,400$. Other parameters are $t=1$, $L=100$, $l=0.2L$, $N_s=200$, and $t'=1.2t$ in (b,c) and $t'=0.7t$ in (d,e).}
\label{figS4}
\end{figure}

We consider a modified SSH Hamiltonian with nonreciprocal and disordered hoppings that cannot be mapped to a Hermitian Hamiltonian through the similarity transformation under OBCs. The modified model contains an additional intercell hopping term with the strength $t''$, as depicted in Fig. \ref{figS4}(a). This hopping term preserves the chiral symmetry of the system, however, the similarity transformation is inapplicable when $t''\neq0$ (see a similar example without disorders in Ref. \cite{SYao2018}) and thus the modified model in this case does not has a direct Hermitian counterpart.

Since the chiral symmetry remains when $t''\neq0$, we can still calculate the disordered-averaged winding number $\nu$. In Fig. \ref{figS4}(b), we show $\nu$ as a function of $W$ ($W_1=W$ and $W_2=0$) and $\gamma$ for $t''=0.2t$ ($t'=1.2t$ and $L=100$). We  find that the non-Hermitian enhancement effect on the topological phase exhibits in this case, although the energy spectrum is generally complex since the similarity transformation is inapplicable. An example of the complex energy spectrum for a disorder configuration and $\gamma=1$ is shown in Fig. \ref{figS4}(c). Moreover, we find that the NHTAI phase can exist without the similarity transformation when $t''\neq0$, with an example for $t''=0.1t$ ($t'=0.7t$ and $L=100$) shown in Fig. \ref{figS4}(c). To show the parameter regime for the NHTAI phase more clearly, we plot $\nu$ as a function of $W$ for $\gamma=0.5t$ and $L=100,200,400$ in Fig. \ref{figS4}(e). We note that the enhancement effect and the NHTAI phase will be broken for large $t''$.

\subsection{4. Results for the case of random non-Hermiticity}

In this section, we consider the effect of random non-Hermiticity on the topological phenomena described in the main text. The hopping terms can be rewritten as
\begin{equation}
m_j=t+W_1\omega_{j}, ~t_j^{(l)}=t'+W_2\omega'_{j}, ~t_j^{(r)}=t_j^{(l)}+t\gamma_j.
\end{equation}
Here $t=1$ is set as the energy unit, $\omega_{j}$ and $\omega'_{j}$ are random numbers chosen uniformly in the range $[-1,1]$ with the disorder strengths $W_1$ and $W_2$, and the nonreciprocal parameters for the $j$-th lattice cell $\gamma_j$ are independent random numbers satisfying the Gaussian distribution $N(0,\sigma^2_{\gamma})$ with the standard deviation $\sigma_{\gamma}$. For simplicity, we focus on the case of $W_2=0$ and set $W_1=W$.

\begin{figure}[t]
\centering
\includegraphics[width=0.45\textwidth]{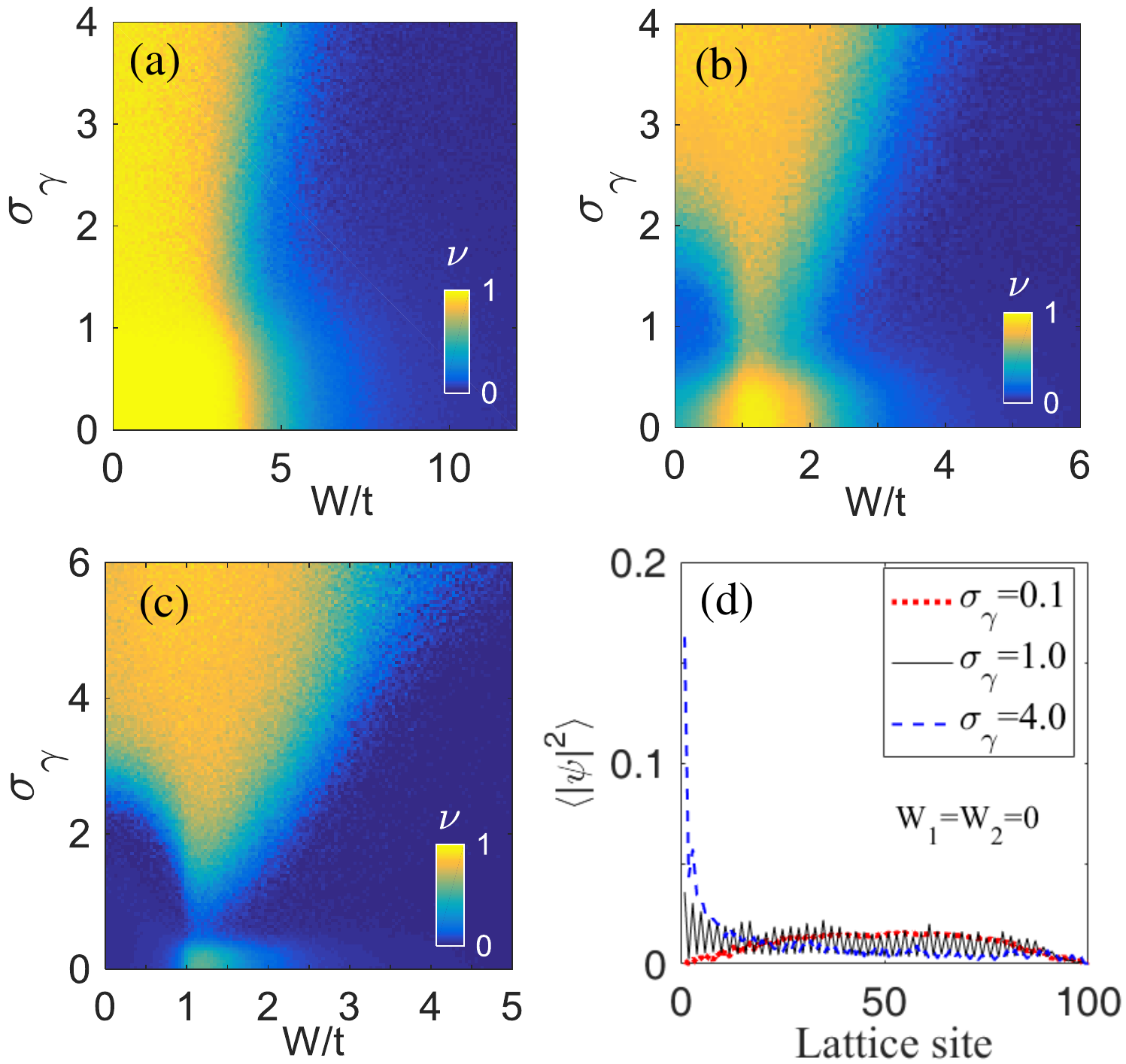}
\caption{(Color online) (a,b,c) The disorder-averaged winding number $\nu$ as a function of the disorder strength $W$ and the standard deviation of random non-Hermiticity $\sigma_{\gamma}$ for $t'=2t$, $t'=t$, and $t'=0.7t$, respectively. (d) The averaged density distribution of the $L/2$-th eigenstate for $\sigma_{\gamma}=0.1,1,4$ with fixed $W_1=W_2=0$ and $t'=0.7t$. Other parameters are $t=1$, $L=100$, $l=0.2L$, $W_1=W$, $W_2=0$, and $N_s=1000$.}
\label{figS5}
\end{figure}

In the presence of hopping disorder and random nonreciprocal parameter, we numerically find that this kind of non-Hermiticity cannot enhance the topological phase in the general case when $t'>t$ , with an typical example shown in Fig. \ref{figS5}(a) for $t'=2t$. However, the random non-Hermiticity still can induce and enhance the topological phase from the critical case of $t'=t$, as shown in Fig. \ref{figS5}(b), similar as the results shown in Fig. 2(e) in the main text. Moreover, the NHTAI phase still exists in under this condition by the combination of hopping disorders and random non-Hermiticity from the trivial phase in the clean and Hermitian limit ($t'<t$), as shown in Fig. \ref{figS5}(c) for $t'=0.7t$. Notably, under the random non-Hermiticity $\gamma_j$ with the zero mean value, the non-Hermitian skin effect of the bulk states in the open chain is generally broken after averaging over many configurations. As shown in Fig. \ref{figS5}(d), we plot the averaged density distribution of the $L/2$-th eigenstate for $\sigma_{\gamma}=0.1,1,4$ with fixed $W_1=W_2=0$ and $t'=0.7t$. It is clear that when $\sigma_{\gamma}=0.1,1$, this bulk state is extended in the lattice under OBCs, which indicates the breaking of the non-Hermitian skin effect. Other extended bulk states exhibit similar density distributions. When $\sigma_{\gamma}=4$, the $L/2$-th ($L/2+1$-th) eigenstate becomes a zero-energy mode localized near the left (right) edge of the system that is in the topological phase.

\subsection{5. Disordered SSH model with gain and loss}

\begin{figure}[t]
\centering
\includegraphics[width=0.45\textwidth]{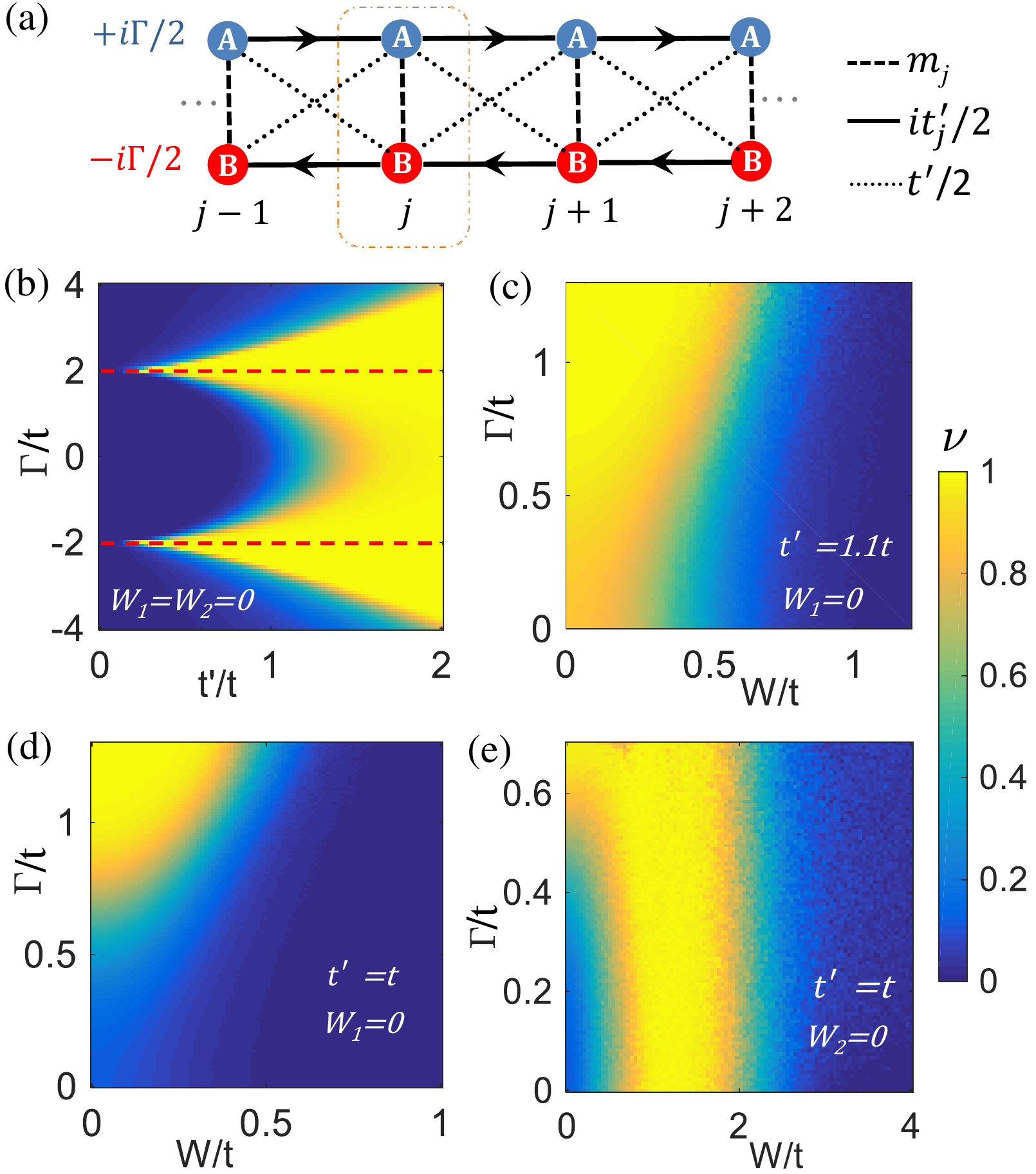}
\caption{(Color online) (a) Sketch of the disordered SSH model with non-Hermitian gain and loss, denoted by $\pm i\Gamma/2$. The dotted box is the unit cell, and $m_j$, $it'_j/2$, and $t'/2$ denote the corresponding hopping strengths. (b) The winding number $\nu$ as a function of $t'$ and $\Gamma$ for $W_1=W_2=0$ and $L=24$. The red dashed lines denote the $PT$ symmetry breaking threshold. (c) $\nu$ as a function of $W$ and $\Gamma$ for $t'=1.1t$, $W_2=W$ and $W_1=0$. (d,e) $\nu$ as a function of $W$ and $\Gamma$ for $t'=t$, $W_1=W$ (with $W_2=0$) and $W_2=W$ (with $W_1=0$), respectively. Other parameters are $t=1$, $L=100$, $l=0.2L$ and $N_s=200$.}
\label{figS6}
\end{figure}

We further consider another disordered SSH-type model with non-Hermitian gain and loss, as depicted in Fig. \ref{figS6}(a). The model Hamiltonian is given by
\begin{align}
H_{2}&=\sum_{j}(m_ja_{j}^{\dag}b_{j}+h.c.)+\sum_{j}\frac{i\Gamma}{2}(a_{j}^{\dag}a_{j}-b_{j}^{\dag}b_{j}),\\ \nonumber
&+\sum_{j}[\frac{it'_{j}}{2}(a_{j+1}^{\dag}a_{j}-b_{j+1}^{\dag}b_{j})+\frac{t'}{2}a_{j+1}^{\dag}b_{j}+h.c.], \nonumber
\label{Ham2}
\end{align}
with the gain-and-loss parameter $\Gamma$ and the random hopping strengths
\begin{equation}
m_j=t+W_1\omega_{j}, ~it'_j=it'+W_2\omega'_{j}.
\end{equation}
Here $\omega_{j}$ and $\omega'_{j}$ are independent random real numbers chosen uniformly in the range $[-1,1]$, $W_1$ and $W_2$ are the corresponding disorder strengths. In the clean limit $W_1=W_2=0$, the corresponding Bloch Hamiltonian in momentum ($k$) space is given by $\mathcal{H}_2(k)=(t+t^{\prime} \cos k)\sigma_x+(t^{\prime} \sin k-i\Gamma/2)\sigma_z$, which has been proposed in Ref. \cite{Lee2016}. This model is also studied in Ref. \cite{SYao2018} under a (pseudo-)spin rotation $\sigma_z\rightarrow\sigma_y$, which corresponds to the unitary transformation $\mathcal{U}\mathcal{H}_2(k)\mathcal{U}^{-1}=\mathcal{H}'_2(k)$, with $\mathcal{U}=(i\sigma_x+\mathbb{I}_2)/\sqrt{2}$ and $\mathbb{I}_2$ as the $2\times2$ identity matrix. The Bloch Hamiltonian $\mathcal{H}_2(k)$ has the chiral (sublattice) symmetry $\mathcal{C}\mathcal{H}_2(k)\mathcal{C}^{-1}=-\mathcal{H}_2(k)$ and the parity-time ($\mathcal{PT}$) symmetry $\mathcal{P T}\mathcal{H}_2(k)(\mathcal{P T})^{-1}=\mathcal{H}_2(-k)$, where the chiral, parity and time reversal operators are respectively defined by $\mathcal{C}=\sigma_y$, $\mathcal{P}=\sigma_x$ and $\mathcal{T}=\mathcal{K}$ with the complex conjugation $\mathcal{K}$.

These symmetries can also be introduced for the system under OBCs. The real-space Hamiltonian $H_2$ under OBCs has the chiral symmetry $CH_2C^{-1}=-H_2$, where the chiral operator is $C=\sigma_y\otimes \mathbb{I}_N$ with $\mathbb{I}_N$ the $N \times N$ identity matrix for the system of $N$ lattice cells. The parity-time ($PT$) symmetry for $H_2$ under OBCs is defined by $PTH_2(PT)^{-1}=H_2$, with the parity operator $P=\sigma_x\otimes \mathbb{I}_N$ and the time-reversal operator $T=\mathcal{K}\otimes \mathbb{I}_N$. In the presence of disorders, the chiral symmetry preserves and thus the (disorder-averaged) open-bulk winding number in real space can be calculated in this model (by rewriting the $Q_s$ matrix with two chiral-symmetric parts defined by $C=\sigma_y\otimes \mathbb{I}_N$ here).

For the clean case of $W_1=W_2=0$, the $PT$ symmetry remains unbroken in the system under OBCs before $|\Gamma|$ reaches the threshold $|\Gamma|=2t$. In this case, we find that the increase of the non-Hermitian gain-and-loss parameter $|\Gamma|$ can enlarge the parameter regime of the topological phase with $\nu=1$, as shown in Fig. \ref{figS6}(b). In particular, the topological phase can be induced from an initially Hermitian trivial phase with $t'/t<1$ and $\Gamma=0$ by turning on and increasing the gain-and-loss non-Hermiticity. The induced topological phase will return to the trivial phase for large $|\Gamma|$ in the $PT$ broken region. Note that the results preserve under the transformation $\Gamma\rightarrow-\Gamma$. The enhancement of the topological phase by the gain and loss in the presence of the $PT$ symmetry can also be understood by a similarity transformation as the energy spectrum is real under this symmetry. We apply a unitary transformation on the Hamiltonian $H_2$ under OBCs: $UH_2U^{-1}=H'_2$, where the transformation operator $$U=\frac{1}{\sqrt{2}}\left(
                                                                                                                               \begin{array}{cc}
                                                                                                                                 1 & i\\
                                                                                                                                 i & 1 \\
                                                                                                                               \end{array}
                                                                                                                             \right)
\otimes \mathbb{I}_N.$$
Then the real-space Hamiltonian $H'_2$ describes an open SSH chain with nonreciprocal intracell hopping terms $t\pm\Gamma/2$ \cite{SYao2018}. When $|\Gamma|<2t$, there exists a similarity transformation mapping the non-Hermitian Hamiltonian $H'_2$ onto a Hermitian counterpart $\tilde{H}'_2$ with intracell hopping strength $\tilde{t}=\sqrt{t^2-\Gamma^2/4}$ and intercell hopping strength $\tilde{t'}=t'$. Thus, the increase of $|\Gamma|$ effectively increases the ratio $\tilde{t'}/\tilde{t}$ up to $|\Gamma|=2t$, which leads to the enhancement of the topological phase with $\tilde{t'}/\tilde{t}>1$. Note that here $t=1$ as the energy unit and the topological transition point is $\tilde{t'}/\tilde{t}=1$.

In the presence of hopping disorders with $W_2=W$ and $W_1=0$ under the $PT$ symmetry, the non-Hermitian gain-and-loss still enhances the topological phase in the open chain with $t'>t$ against disorders, as shown in Fig. \ref{figS6}(c). In addition, we find that the gain-and-loss can induce and enhance the topological phase from the critical case of $t=t'$, as shown in Fig. \ref{figS6}(d). When $W_1\neq0$, the $PT$ symmetry can be spontaneously broken in certain disorder configurations with large $W_1$, and then the non-Hermitian enhancement effect in the region of $t'>t$ may disappear (not shown here). However, the NHTAI phase can still be induced from the critical case in this SSH model by the combination of disorder ($W_1=W$ and $W_2=0$) and gain-and-loss non-Hermiticity, as shown in Fig. \ref{figS6}(e). We notice that the proposed disordered SSH model with non-Hermitian gain and loss is realizable in some artificial systems, such as ultracold atoms, photonic crystals, and electronic circuits.

\end{document}